# Real-time dual frequency comb spectroscopy in the near infrared

**F. Zhu,[1*] T. Mohamed,[1,2,3] J. Strohaber,[1] A. A. Kolomenskii,[1] Th. Udem,[4] and H. A. Schuessler[1,2]**

[1] *Department of Physics and Astronomy, Texas A&M University, College Station, TX 77843-4242, USA*
[2] *Science Department, Texas A&M University at Qatar, Doha 23874, Qatar*
[3] *Physics Department, Faculty of Science, Beni-Suef University, Beni Suef 62511, Egypt*
[4] *Max-Planck-Institut für Quantenoptik, 85748 Garching, Germany*
**Corresponding author: zhuf@physics.tamu.edu*

We use two femtosecond Erbium-doped fiber lasers with slightly different repetition rates to perform a modern type of Fourier transform spectroscopy without moving parts. The measurements are done in real time, and it takes less than 50 μs. We work with somewhat different spectral outputs from two Erbium-doped fiber lasers and employ spectral filtering based on a *2f-2f* grating setup to select the common spectral region of interest, thereby increasing the signal-to-noise ratio. The interferogram is taken with a 20 cm long gas cell, containing a mixture of acetylene and air at atmospheric pressure, and is fast-Fourier-transformed to obtain the broadband spectral fingerprint of the gas.

Ever since the femtosecond frequency comb was invented in the late 1990s [1], there have been ongoing revolutions in the field of spectroscopy. Many methods have been and are being developed to utilize the regular comb structure of millions of laser modes in spectroscopy ranging from the XUV to the mid IR [2-5]. Among these, dual frequency comb spectroscopy (DFCS) emerges as a promising, highly sensitive, superior fast spectroscopy with high resolution to complement the traditional Fourier Transform Spectroscopy (FTS) [2, 6-12]. Especially in the near and mid IR, a plethora of greenhouse and other gases have molecular fingerprint spectra that can be studied with DFCS based mainly on Er- or Yb-doped fiber lasers and their wavelength ranges extended by optical parametric oscillation process [13], supercontinuum, or difference frequency generation [14].

DFCS uses two femtosecond frequency combs with slightly different repetition rates. In the time domain, pulse pairs arrive at a photo detector (PD) with a linear increasing time delay. Each time a pulse pair overlaps in time, like the zero optical path difference in FTS, the central burst of an interferogram is formed. Subsequent pairs of pulses impinge on the PD with varying delay, analogous to the delay introduced by FTS, except no moving mirror parts are needed. As a result, the PD records an interferogram formed by many pulse pairs of various delay. Because pulse pairs repeatedly move through each other, a new interferogram starts to form as soon as the previous is completed. In the frequency domain, comb lines of one source beat with the same order comb lines of the other source, and the optical frequency information is down converted to the radio frequency range, which is the Fourier transform of the interferogram. After frequency up conversion and calibration, the optical spectrum is recovered.

Compared to traditional FTS, DFCS is extremely fast, and an interferogram can be recorded in less than 25 μs [9]. A fast oscilloscope can display an interferogram and the broadband spectrum by Fast Fourier Tranform (FFT) in real time. If the repetition rates and carrier envelope offset (CEO) frequencies of both combs are well stabilized, several seconds of the interference signal can be transformed to a spectrum with high resolution at high signal-to-noise ratio (SNR) or be reconstructed to get the free induction decay (FID) response in the time domain [6, 8, 10]. Because DFCS requires two comb sources, most of the applications employed two identical ones. In this letter, we present results of real-time DFCS with two different types of comb sources and a novel spectral filtering method to increase the SNR.

Our experimental setup is depicted in Fig. 1. Fiber comb 1 (Menlo Systems, M-comb) is a femtosecond Er-doped fiber oscillator with an output power of ~25 mW. Fiber comb 2 (Menlo Systems, M-fiber) is a similar oscillator with an Er-doped fiber amplifier. The amplified pulse is coupled into a photonic crystal fiber to generate a red-shifted Raman soliton through asymmetric broadening. The blue part of the spectrum acquires oscillations due to self-phase modulation. The total output power is ~500 mW. Both fiber combs operate at repetition rates locked to frequencies of ~250 MHz, and their CEO frequencies are not stabilized. The autocorrelation traces and spectra of both comb sources are depicted in the insets of Fig. 1. Slightly asymmetric traces, in particular of the spectrally broader fiber comb 2, are due to the dispersion mismatch for the two arms of our home build interferometric autocorrelator. The pulse durations are about 67 fs and 86 fs for fiber combs 1 and 2 respectively. The laser beam from fiber comb 2 passes through a 20 cm long gas cell filled with a $C_2H_2$ and air mixture and overlaps spatially with the beam from fiber comb 1 with a polarizing beam splitter (PBS) cube.

Typically, the molecular absorption spectrum of interest covers only a small portion of the available spectrum, and the extraneous portion only increases noise, since it is accumulated over the whole spectrum. To reduce noise in earlier work, tunable narrow-bandwidth fiber Bragg-gratings or bandpass filters were used, and the broadband spectrum was stitched together [6, 10]. This method provides high SNR and high resolution, but requires more time; therefore, it is not ideal for real-time spectroscopy. We employ a grating based spectral filter in a *2f-2f* configuration, which is simple, versatile and compatible with real-time measurements. The filter is easily tuned to cover the whole spectral region of interest, therefore stitching is not required. To achieve this, the overlapped laser beams are directed to a grating. The first order diffraction is focused by a lens of focal length $f$ positioned at a distance of $2f$ away from the grating. To select the spectral region of interest, two blades on translational stages are placed in the focal plane of the lens as an adjustable slit. The PD is placed $2f$ away from the lens to detect the beams recombined within the selected spectral slice [15].

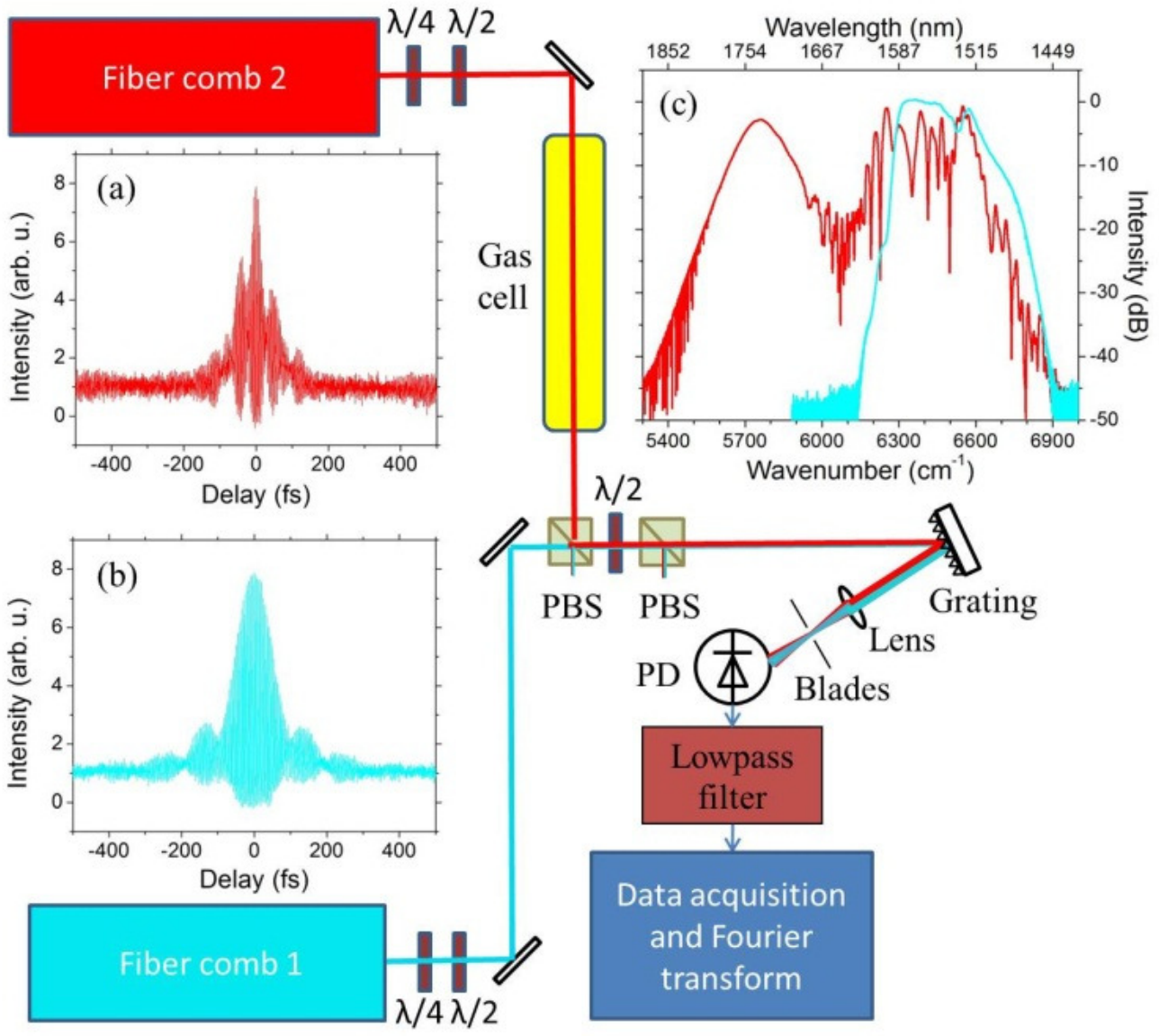

Fig. 1. (Color online) Experimental setup includes two fiber lasers, waveplates and polarizers allowing to independently adjust the powers of the beams, a $2f$-$2f$ spectral filter and a photodetector with electronics. Interferometric autocorrelation traces are shown in inset (a) for fiber comb 2 (red) and (b) for fiber comb 1 (cyan) and inset (c) shows their respective spectra.

To acquire spectral data, the difference between the repetition rates of the two fiber combs is set at 1807 Hz. This value is sufficiently low (<2.6 kHz) to avoid comb modes beating at frequencies larger than half of the repetition rate, while it is large enough to allow for a fast recording time.

To reduce aliasing, a low pass filter (DC-140 MHz) is used to filter out frequencies higher than half of the repetition rate. Interferograms are recorded with a Tektronix DPO 3054 oscilloscope at a sampling rate of 250 MHz with 11 bits resolution for 40 μs. The PD (Thorlabs, PDA10CF) has a bandwidth of 150 MHz, and shows saturation behavior above 3.0 V. Therefore the total power impinging on the PD must be carefully adjusted to avoid saturation. The specified root mean square noise is 1.4 mV. Assuming that the typical amplitude of the interferogram is about 800 mV, the PD sets the dynamic range limit of the system to about 570 (55 dB in signal power).

A typical interferogram is shown in Fig. 2. The center burst of the interferogram corresponds to the temporal overlap of the pulses from the two combs. On the side of the center burst, the FID tail can be seen.

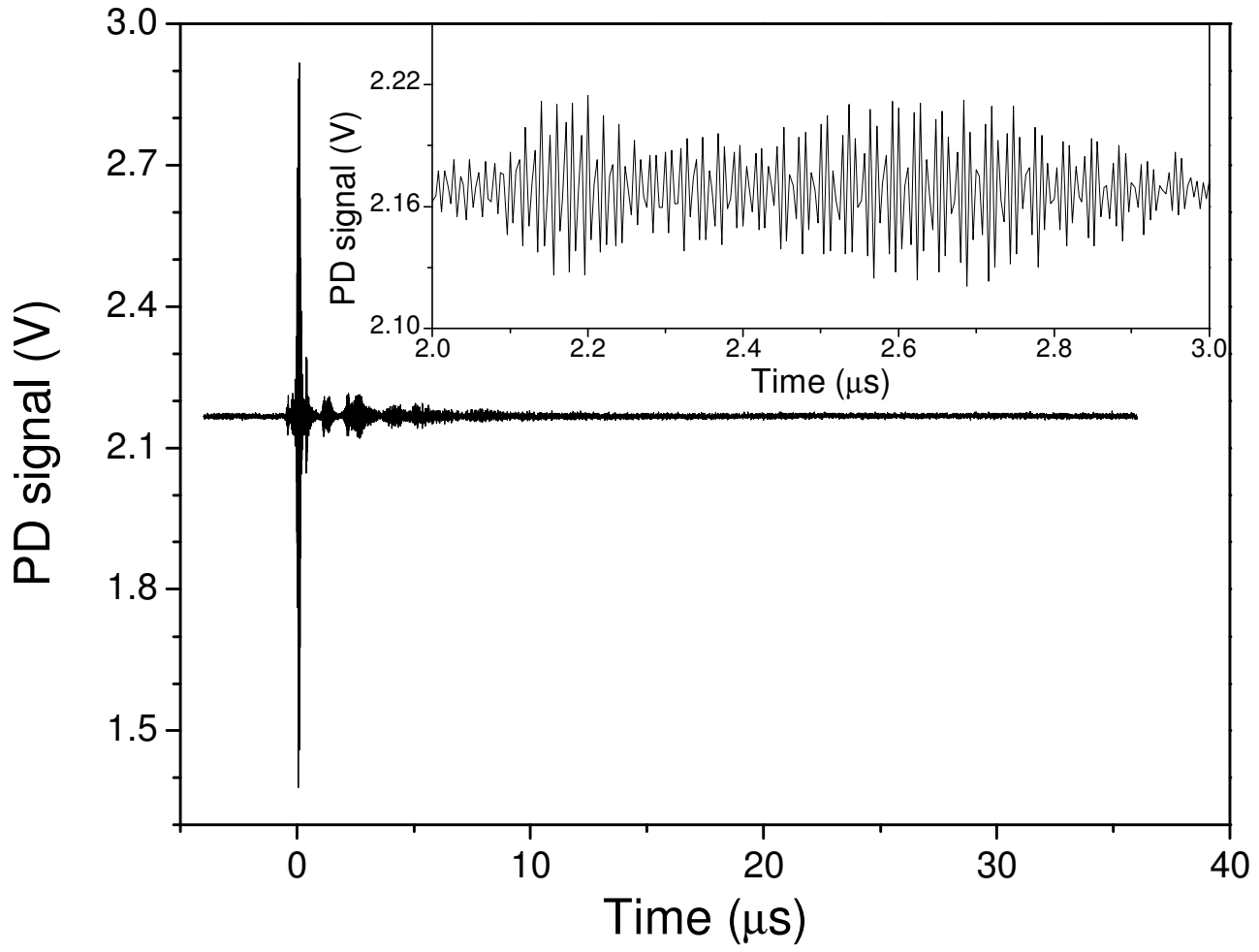

Fig. 2. A typical interferogram. Inset: expanded view of a 1μs long portion of the FID tail.

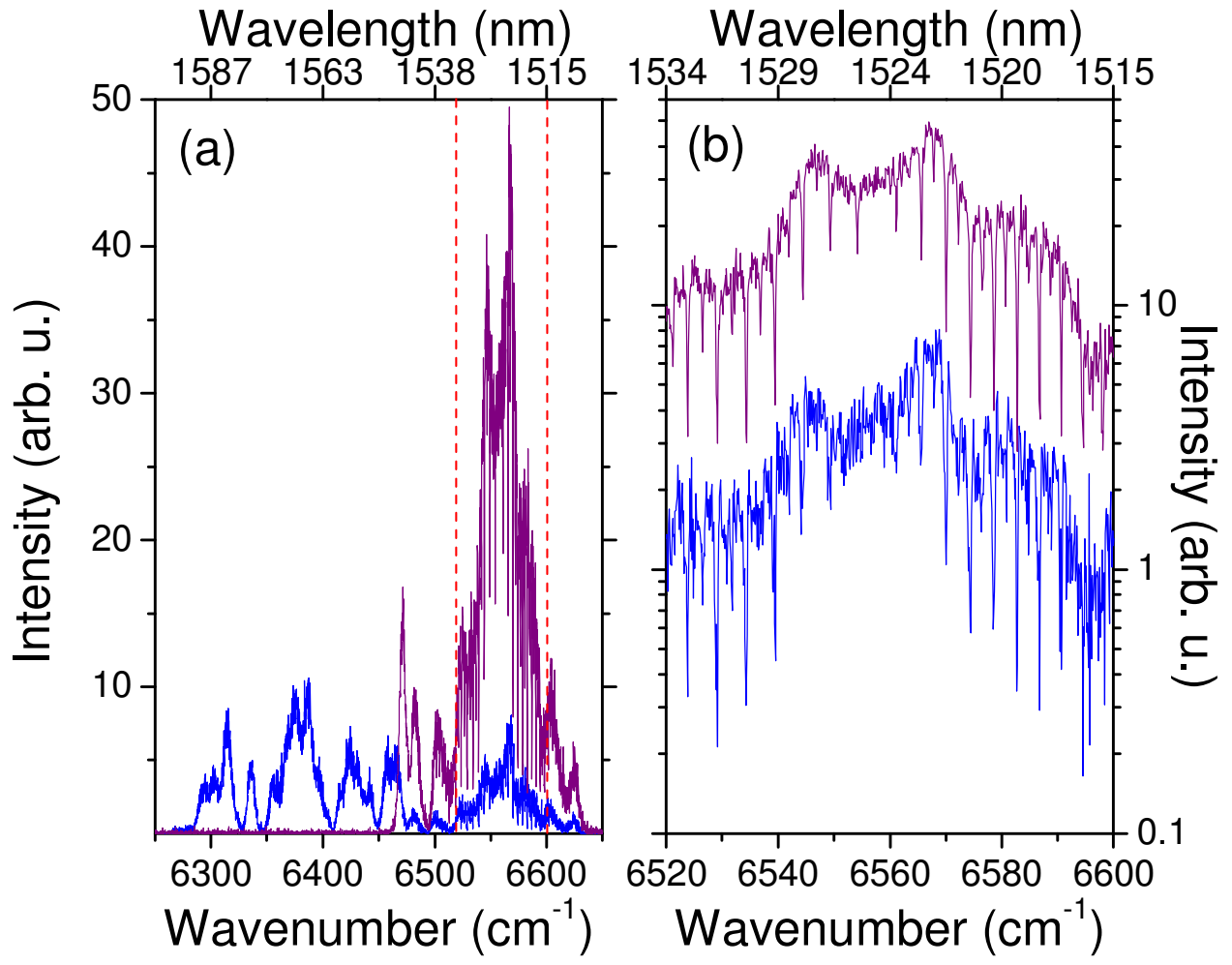

Fig. 3. (Color online) Spectra retrieved from interferograms: (a) with spectral filtering between 6280~6630 cm$^{-1}$ (blue), and between 6460~6630 cm$^{-1}$ (maroon) corresponding to the spectral fingerprint interval of acetylene are shown. A six-fold beat signal increase is achievable when using narrower spectral filtering (details in the text). (b) A magnified view of the spectral region indicated by dashed vertical lines in panel (a) presented in logarithmic scale. The spectrum with narrower spectral filtering (maroon) shows clearer spectral features with less noise.

Figure 3 shows spectra retrieved by FFT of the corresponding single interferogram recorded for 40 μs. Since in our real-time spectroscopy the CEO frequencies are not known, recordings of the spectra are calibrated against two prominent, well-separated spectral peaks of $C_2H_2$ from the HITRAN database [16]. In Fig. 3 (a), the spectral measurements by DFCS are shown: (1) with spectral filtering between 6280~6630 cm$^{-1}$ (blue), corresponding to the bandwidth of fiber comb 1, and (2) between 6460~6630 cm$^{-1}$, corresponding to the spectral fingerprint interval of acetylene (maroon). The observed larger-scale spectral variations are due to self-phase modulation of the fiber comb 2 (Fig. 1). For both measurements the powers of the two beams reaching the PD were made equal and held constant at a level below saturation of the PD. Similar spectral bandwidths and powers with filtering result in nearly equal amplitudes of the spectral components of the two combs. For measurement (2), when selecting only the fingerprint spectral region, a ~6 times stronger spectral signal is obtained, because spectral filtering allows increasing the power in the desired spectral region. The inset shows the central portion of the spectra on a logarithmic scale, demonstrating a measured ~3 fold improvement of the SNR when narrower filtering was implemented.

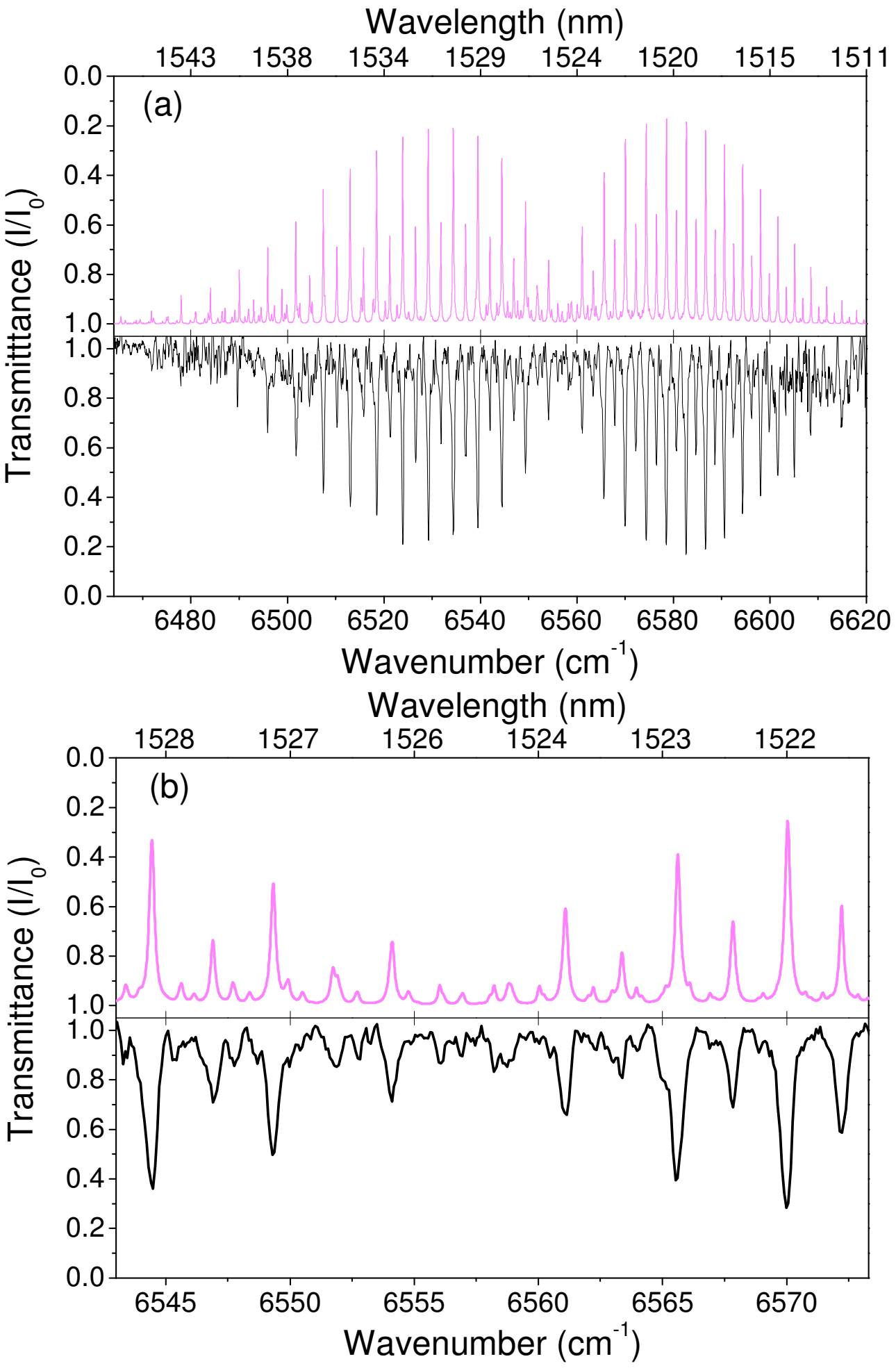

Fig. 4. (Color online) Normalized DFCS spectra of acetylene (black) compared to spectra calculated with the HITRAN data base (pink, inverted for clarity): (a) in the broad range between 6466~6620 $cm^{-1}$, (b) expanded view in the interval between 6543~6573 $cm^{-1}$.

By narrowing the spectral region to the interval of interest the difference between the repetition rates could be increased to 2426 Hz, which also sped up the data acquisition rate. To improve the SNR of the measurements prior to FFT, we perform averaging of four consecutive interferograms, so that the total acquisition time is 1.65 ms. To extract the normalized spectrum, we first take the absorption spectrum with the $C_2H_2$ (~7%) and air (~93%) mixture at atmospheric pressure in room temperature. Then we use dry air to purge the gas cell and measure the reference spectrum under the same conditions. The spectral transmittance is shown in Fig. 4. Comparison with calculations using the HITRAN data base [16] shows good agreement. To measure the quality of our normalized spectra recorded from 6466 to 6620 $cm^{-1}$ with a resolution of 0.086 $cm^{-1}$, we determine the number of spectral elements to be 1790. By evaluating the root mean square noise level from 6466 to 6471 $cm^{-1}$, the SNR of the most intense line is about 50. Because of the reduced and unequal comb power in some portions of spectrum, the SNRs of these areas are less. Using the quality factor introduced by Ref.11, which is the product of the number of spectral elements and the SNR normalized by the square root of the total acquisition time counting both absorption and reference scans, we obtain an experimental quality factor of about $1.6\times10^6$ $Hz^{1/2}$, which is similar to the values of the previous works [7, 9, 10, 11].

Because of the pulse to pulse jitter, the intensity profile fluctuates, which results in noise of different parts of the normalized spectrum as the absorption and reference spectra are not recorded simultaneously. This can be improved once the CEO frequencies of both combs are stabilized, so that interferograms can be recorded and averaged over a longer time. In addition an adaptive sampling scheme is being developed to achieve high-fidelity real-time DFCS [12].

Although real-time DFCS needs to overcome many technical challenges (such as developing a low noise PD with large bandwidth and high dynamic range, comb sources with low intensity noise and spectrally flat, phase-coherent broad spectra) the fast and relatively simple scheme is promising for the rapid identification of many molecular species simultaneously in real-time over a wide spectral range. The sensitivity can be further improved with such techniques as cavity [9] or multipass signal enhancement to characterize minute trace amounts of molecules.

We thank Menlo Systems for the generous loan of the M-comb. This work is funded by the Robert A. Welch Foundation grant No. A1546, and the NSF grant No. 1058510.

## References

1. Th. Udem, R. Holzwarth, and T. W. Hänsch, Optical frequency metrology, Nature **416**, 233-237 (2002).
2. F. Keilmann, C. Gohle, and R. Holzwarth, Time-domain mid-infrared frequency-comb spectrometer, Opt. Lett. **29,** 1542-1544 (2004).
3. M. J. Thorpe, D. D. Hudson, K. D. Moll, J. Lasri, and J. Ye, Cavity-ring down molecular spectroscopy based on an optical frequency comb at 1.45-1.65 μm. Opt. Lett. **32**, 307-309 (2007).
4. S. A. Diddams, L. Hollberg, and V. Mbele, Molecular fingerprinting with the resolved modes of a femtosecond laser frequency comb, Nature **445**, 627-630 (2007).
5. A. Cignöz, D. C. Yost, T. K. Allison, A. Ruehl, M. E. Fermann, I. Hartl, and J. Ye, Direct frequency comb spectroscopy in the extreme ultraviolet, Nature **482**, 68-71 (2012).
6. I. Coddington, W. C. Swann, and N. R. Newbury, Coherent multiheterodyne spectroscopy using stabilized optical frequency combs, Phys. Rev. Lett. **100**, 013902 (2008).

7. P. Giaccari, J. D. Deschenes, P. Saucier, J. Genest, and P. Tremblay, Active Fourier-transform spectroscopy combining the direct RF beating of two fiber-based mode-locked lasers with a novel referencing method, Opt. Exp. **16**, 4347-4365 (2008).
8. I. Coddington, W. C. Swann, and N. R. Newbury, Time-domain spectroscopy of molecular free-induction decay in the infrared, Opt. Lett. **35**, 1395-1397 (2010).
9. B. Bernhardt, A. Ozawa, P. Jacquet, M. Jacqey, Y. Kobayashi, Th. Udem, R. Holzwarth, G. Guelachvili, T. W. Hänsch, and N. Picque, Cavity-enhanced dual-comb spectroscopy, Nature Photonics, **4**, 55-57 (2010).
10. I. Coddington, W. C. Swann, and N. R. Newbury, Coherent dual-comb spectroscopy at high signal-to-noise ratio, Phys. Rev. A **82**, 043817 (2010).
11. N. R. Newbury, I. Coddington, and W. Swann, Sensitivity of coherent dual-comb spectroscopy, Opt. Exp. **18**, 7929-7945 (2011).
12. T. Ideguchi, A. Poisson, G. Guelachvili, N. Pique, and T. W. Hänsch, Adaptive real-time dual-comb spectroscopy, arXiv: 1201.4177 (2012).
13. Z. Zhang, C. Gu, J. Sun, C. Wang, T. Gardiner, and D. T. Reid, Asynchronous midinfrared ultrafast optical parametric oscillator for dual-comb spectroscopy, Opt. Lett. **37**, 187-189 (2012).
14. T. W. Neely, T. A. Johnson, and S. A. Diddams, High-power broadband laser source tunable from 3.0 μm to 4.4 μm based on a femtosecond Yb:fiber oscillator, Opt. Lett. **36**, 4020-4022 (2011).
15. I. G. Mariyenko, J. Strohaber, and C. J. G. J. Uiterwaal, Creation of optical vortices in femtosecond pulses, Opt. Exp. **13**, 7599-7608 (2005).
16. L. S. Rothman, et al. The HITRAN 2008 molecular spectroscopic database, J. Quant. Spectrosc. Radiat. Transfer, **110**, 533-572 (2009).